\documentclass[preprint,aps]{revtex4}

\usepackage{graphicx}

\newcommand{\bubun}[2]{\frac{\partial #1}{\partial #2}}

\newcommand{\infinity}{\infty}

\begin{document}

\preprint{ }

\title{Evolution Equation of Phenotype Distribution: General Formulation and Application
to Error Catastrophe} \author{Katsuhiko Sato$^1$ and Kunihiko Kaneko$^{1,2}$} \address{
$^1$ Complex Systems Biology Project, ERATO JST} \address{ $^2$ Department of Pure and
Applied Sciences, University of Tokyo, 3-8-1 Komaba, Meguro-ku, Tokyo 153-8902, Japan }
\email{sato@complex.c.u-tokyo.ac.jp; kaneko@complex.c.u-tokyo.ac.jp} \date{\today}

\begin{abstract} 
An equation describing the evolution of phenotypic distribution is derived using methods
developed in statistical physics.  The equation is solved by using the singular
perturbation method, and assuming that the number of bases in the genetic sequence is
large.  Applying the equation to the mutation-selection model by Eigen provides the
critical mutation rate for the error catastrophe.  Phenotypic fluctuation of clones
(individuals sharing the same gene) is introduced into this evolution equation.  With this
formalism, it is found that the critical mutation rate is sometimes increased by the
phenotypic fluctuations, i.e., noise can enhance robustness of a fitted state to mutation. 
Our formalism is systematic and general, while approximations to derive more tractable
evolution equations are also discussed.
\end{abstract}

\pacs{ }


\maketitle

\section{Introduction}

For decades, quantitative studies of evolution in laboratories have used bacteria and
other microorganisms\cite{Lenski,Lenski-Rose,Kishony}.  Changes in phenotypes, such as
enzyme activity and gene expressions introduced by mutations in genes, are measured along
with the changes in their population distribution in phenotypes
\cite{Bactreia-many-generations, Dekel-Alon,Kashiwagi-Noumachi,Ito}.  Following
such experimental advances, it is important to analyze the evolution equation of
population distribution of concerned genotypes and phenotypes.

In general, fitness for reproduction is given by a phenotype, not directly by a genetic
sequence. Here, we consider evolution in a fixed environment, so that the fitness is given
as a fixed function of the phenotype.  A phenotype is determined by mapping a genetic
sequence. This phenotype is typically represented by a continuous (scalar) variable, such
as enzyme activity, protein abundances, and body size. For studying the evolution of a
phenotype, it is essential to establish a description of the distribution function for a
continuous phenotypic variable, where the fitness for survival, given as a function of
such a continuous variable, determines population distribution changes over generations.

However, since a gene is originally encoded on a base sequence (such as AGCTGCTT in DNA),
it is represented by a symbol sequence of a large number of discrete elements. Mutation in
a sequence is not originally represented by a continuous change.  Since the fitness is
given as a function of phenotype, we need to map base sequences of a large number of
elements onto a continuous phenotypic variable $x$, where the fitness is represented as a
function of $x$, instead of the base sequence itself.  A theoretical technique and careful
analysis are needed to project a discrete symbol sequence onto a continuous
variable. 

Mutation in a nucleotide sequence is random, and is represented by a stochastic process.
Thus, a method of deriving a diffusion equation from a random walk is often
applied. However, the selection process depends on the phenotype. If a phenotype is given
as a function of a sequence, the fitness is represented by a continuous variable mapped
from a base sequence. Since the population changes through the selection of fitness, the
distribution of the phenotype changes accordingly. If the mapping to the phenotype
variable is represented properly, the evolutionary process will be described by the
dynamics of the distribution of the variable, akin to a Fokker-Planck equation.

In fact, there have been several approaches to representing the gene with a continuous
variable \cite{footnote}
.  Kimura\cite{Kimura2} developed the population distribution of a continuous fitness.
Also, for certain conditions, a Fokker-Planck type equation has been analyzed by
Levine\cite{Levine}. Generalizing these studies provides a systematic derivation of an
equation describing the evolution of the distribution of the phenotypic variable. We adopt
selection-mutation models describing the molecular biological evolution discussed by
Eigen\cite{Eigen}, Kauffman\cite{kauffmann-book}, and others, and take a continuum limit
assuming that the number of bases $N$ in the genetic sequence is large, and derive the
evolution equation systematically in terms of the expansion of $1/N$.

In particular, we refer to Eigen's equation\cite{Eigen}, originally introduced for the
evolution of RNA, where the fitness is given as a function of a sequence. Mutation into a
sequence is formulated by a master equation, which is transformed to a diffusion-like
equation.  With this representation, population dynamics over a large number of species is
reduced to one simple integro-differential equation with one variable. Although the
equation obtained is a non-linear equation for the distribution, we can adopt techniques
developed in the analysis of the (linear) Fokker-Planck equation, such as the
eigenfunction expansion and perturbation methods.

So far, we have assumed a fixed, unique mapping from a genotype to a phenotype.  However,
there are phenotypic fluctuations in individuals sharing the same genotype, which has
recently been measured quantitatively as a stochastic gene expression
\cite{Koshland,Elowitz,Kaern-Collins,Collins,Furusawa,Ueda,noise-review}.  Relevance of
such fluctuations to evolution has also been
discussed\cite{SatoPNAS,kaneko-book,KKFurusawaJTB,Ancel}.  In this case, mapping from a gene
gives the average of the phenotype, but phenotype of each individual fluctuates around the
average.  In the second part of the present paper, we introduce this isogenic phenotypic
fluctuation into our evolution equation.  Indeed, our framework of Fokker-Planck type
equations is fitted to include such fluctuations, so that one can discuss the effect of
isogenic phenotypic fluctuations on the evolution.

The outline of the present paper is as follows: We first establish a sequence model in
section (\ref{31:setup-and-derivation}). For deriving the evolution equation from the
sequence model, we postulate the assumption that the transition probability of phenotype
values is uniquely determined by the original phenotype value.  The assumption may appear
too demanding at a first sight, but we show that it is not unnatural from the viewpoint of
evolutionary biology. In fact, most models studied so far satisfy this postulate.  With
this assumption, we derive a Fokker-Planck type equation of phenotypic distribution using
the Kramers-Moyal expansion method from statistical
physics\cite{vanKampen,KuboMatsuoKitahara}.  We discuss the validity of this expansion
method to derive the equation, also from a biological point of view.

As an example of the application of our formulation, we study the Eigen's model in section
(\ref{20}), and estimate the critical mutation rates at which error catastrophe occurs,
using a singular perturbation method.  In section (\ref{32:discussion}), we discuss the
range of the applicability of our method and discuss possible extensions to it.

Following the formulation and application of the Fokker-Planck type equations for
evolution, we study the effect of isogenic phenotypic fluctuations.  While fluctuation in
the mapping from a genotype to phenotype modifies the fitness function in the equation,
our formulation itself is applicable.  We will also discuss how this fluctuation changes
the conditions for the error catastrophe, by adopting Eigen's model.

For concluding the paper, we discuss generality of our formulation, and the relevance of
isogenic phenotypic fluctuation to evolution.

\section{Derivation of evolution equation}
\label{31:setup-and-derivation}

We consider a population of individuals having a haploid genotype, which is encoded on a
sequence consisting of $N$ sites (consider, for example, DNA or RNA). The gene is
represented by this symbol sequence, which is assigned from a set of numbers, such as
$\{-1,1\}$. This set of numbers is denoted by $S$. By denoting the state value of the
$i$th site by $s_i$ ($ \in S$), the configuration of the sequence is represented by the
ordered set $s=\{s_1,...,s_N\}$.

We assume that a scalar phenotype variable $x$ is assigned for each sequence $s$.  This
mapping from sequence to phenotype is given as function $x(s)$. Examples of the phenotype
include the activity of some enzyme (protein), infection rate of bacteria virus, and
replication rate of RNA.  In general, the function $x(s)$ is a degenerate function, i.e.,
many different sequences are mapped onto the same phenotypic value $x$.

Each sequence is reproduced with rate $A$, which is assumed to depend only on the
phenotypic value $x$, as $A(x)$; this assumption may be justified by choosing the
phenotypic value $x$ to relate to the replication. For example, if a protein concerns with
the metabolism of a replicating cell, its activity may affect the replication rate of the
cell and of the protein itself.

In the replication of the sequence, mutation generally occurs; for simplicity, we consider
only the substitution of $s(i)$. With a given constant mutation rate $\mu$ over all sites
in the sequence, the state $s'_i$ of the daughter sequence is changed from $s_i$ of the
mother sequence, where the value $s'_i$ is assigned from the members of the set $S$ with
an equal probability. We call this type of mutation symmetric mutation~\cite{Baake2}. The
mutation is represented by the transition probability $Q(s \rightarrow s')$, from the
mother $s$ to the daughter sequence $s'$. The probability $Q$ is uniquely determined from
the sequence $s$, the mutation rate $\mu$, and the number of members of $S$.  The setup so
far is essentially the same as adopted by Eigen et al.\cite{Eigen}, where the fitness is
given as a function of the RNA sequence or DNA sequence of virus.

Now, we assume that the transition probability depends only on the phenotypic value $x$,
i.e., the function $Q$ can be written in terms of a probability function $W$, which
depends only on $x$, $W(x \rightarrow x')$, as
\begin{equation} \sum_{s' \in \{ s'|x'=x(s') \}} Q(s \rightarrow
s')=W(x(s) \rightarrow x') .
\label{1} \end{equation}

This assumption may appear too demanding. However, most models of sequence evolution
somehow adopt this assumption. For example, in Eigen's model, fitness is given as a
function of the Hamming distance from a given optimal sequence.  By assigning a phenotype
$x$ as the Hamming distance, the above condition is satisfied (this will be discussed
later). In Kauffman's NK model, if we set $N \gg 1$, $K \gg 1$, and $K/N \ll 1$, this
assumption is also satisfied (see Appendix \ref{29}). For the RNA secondary structure
model\cite{Waterman}, this assumption seems to hold approximately, from statistical
estimates through numerical simulations. Some simulations on a cell model with chemical
reaction networks\cite{Furusawa,Furusawa-KK} also support the assumption. In fact, a
similar assumption has been made in evolution theory with a gene substitution
process\cite{Gillespie,Orr}.

The validity of this assumption in experiments has to be confirmed. Consider a selection
experiment to enhance some function through mutation, such as the evolution of a certain
protein to enhance its activity\cite{Ito}. In this case, the assumption means that the
activity distribution over the mutant proteins is statistically similar as long as they
have the same activity, even though their mother protein sequences are different.

With the above setup, we consider the population of these sequences and their dynamics,
allowing for overlap between generations, by taking a continuous-time
model\cite{Baake2}. We do not consider the death rate of the sequence explicitly since its
consideration introduces only an additional term, as will be shown later. The
time-evolution equation of the probability distribution $\hat{P}(s,t)$ of the sequence $s$
is given by:
\begin{equation} \bubun{\hat{P}(s,t)}{t} = -\bar{A}(t) \hat{P}(s,t) +
\sum_{s'} A(x(s')) Q(s' \rightarrow s) \hat{P}(s',t), \label{3} \end{equation} as
specified by Eigen\cite{Eigen}. Here the quantity $\bar{A}(t)$ is the average fitness of
the population at time $t$, defined by $\bar{A}(t)=\sum_{s} A(x(s)) \hat{P}(s,t)$ and $Q$
is the transition probability satisfying $\sum_{s} Q(s' \rightarrow s)=1$ for any $s'$.

According to the assumption (\ref{1}), eq. (\ref{3}) is transformed into the equation for
$P(x,t)$, which is the probability distribution of the sequences having the phenotypic
value $x$, defined by $P(x,t)=\sum_{ s \in \{s | x = x(s)\}} \hat{P}(s,t)$. The equation
is given by \begin{equation} \bubun{P(x,t)}{t} = - \bar{A}(t) P(x,t) + \sum_{x'} A(x')
W(x' \rightarrow x) P(x',t), \label{4}
\end{equation}
where the function $W$ satisfies
\begin{equation} \sum_{x} W(x' \rightarrow x)=1 \qquad \mbox{for any $x'$,}
\label{2} \end{equation} \noindent as shown.

Since $N$ is sufficiently large, the variable $x$ is regarded as a continuous variable. By
using the Kramers-Moyal expansion\cite{vanKampen,KuboMatsuoKitahara,Haken}, with the help
of property (\ref{2}), we obtain:
\begin{equation}
\bubun{P(x,t)}{t} = (A(x)-\bar{A}(t)) P(x,t) + \sum_{n=1}^{\infinity} \frac{(-1)^n }{n!} 
\bubun{{}^n}{x^n} m_n(x) A(x) P(x,t),
\label{5} \end{equation}
where $m_n(x)$ is the $n$th moment about the value $x$, defined by
$m_n(x)= \int (x'-x)^n W(x \rightarrow x') dx' $.

Let us discuss the conditions for the convergence of expansion (\ref{5}), without
mathematical rigor.  For convergence, it is natural to assume that the function $W(x'
\rightarrow x)$ decays sufficiently fast as $x$ gets far from $x'$, by the definition of
the moment.

Here, the transition $W(x' \rightarrow x)$ is a result of $n$ point mutants of the
original sequence $s'$ for $n=0,1,2,...,N$. Accordingly, we introduce a set of quantities,
$w_n(x(s') \rightarrow x)$, as the fitness distribution of $n$ point mutants of the
original sequence $s'$ (Naturally, $w_0(x(s') \rightarrow x)=\delta(x(s')-x)$, which does
not contribute to the $n$th moment $m_n$ ($n \geq 1$)). Next, we introduce the probability
$p_n$ that a daughter sequence is an $n$ point mutant $(n=0,1,2,...,N)$ from her mother
sequence, which are determined only by the mutation rate $\mu$ and the sequence length
$N$. Indeed, ${p_n}'s$ form a binomial distribution, characterized by $\mu$ and $N$.

In terms of the quantities $w_n$ and $p_n$, we are able to write down the transition
probability $W$ as
\begin{equation}
W(x(s') \rightarrow x)=\sum_{n=0}^{N} p_n w_n(x(s') \rightarrow x).\label{6}
\end{equation}
Now, we discuss if $W(x(s') \rightarrow x)$ decays sufficiently fast with $|x(s')-x|$.
First, we note that the width of the domain, in which $w_n(x(s') \rightarrow x)$ is not
close to zero, increases with $n$ since $n$-point mutants involve increasing number of
changes in the phenotype with larger values of $n$.  Then, to satisfy the condition for
$W(x(s') \rightarrow x)$, at least the single-point-mutant transition $w_1(x(s')
\rightarrow x)$ has to decay sufficiently fast with $|x(s')-x|$. In other words, the
phenotypic value of a single-point mutant $s$ of the mother sequence $s'$ must not vary
much from that of the original sequence, i.e., $|x(s')-x(s)|$ should not be large
(``continuity condition").

In general, the domain $|x-x(s')|$, in which $w_n(x(s') \rightarrow x) \neq 0$, increases
with $n$. On the other hand, the term $p_n$ decreases with $n$ and with the power of
$\mu^n$. Hence, as long as the mutation rate is not large, the contribution of $w_n$ to
$W$ is expected to decay with $n$. Thus, if the continuity condition with regards to a
single-point mutant and a sufficiently low mutation rate are satisfied, the requirement on
$W(x(s') \rightarrow x)$ should be fulfilled. Hence, the convergence of the expansion is
expected.

Following the argument, we further restrict our study to the case with a small mutation
rate $\mu$ such that $\mu N \ll 1$ holds. The transition probability $W$ in eq.  (\ref{6})
is written as
\begin{equation} W(x(s') \rightarrow x) \simeq (1-\mu N)
\delta(x(s')-x) + \mu N w_1(x(s') \rightarrow x),\label{7}
\end{equation}
where we have used the property that ${p_n}'s$ form the binomial distribution
characterized by $\mu$ and $N$. Introducing a new parameter, $\gamma$ ($\gamma=\mu N$),
that gives the average of the number of changed sites at a single-point mutant, and using
the transition probability (\ref{7}), we obtain
\begin{equation} \bubun{P(x,t)}{t} =
(A(x)-\bar{A}(t)) P(x,t) + \gamma \sum_{n=1}^{\infinity} \frac{(-1)^n
} {n!} \bubun{{}^n}{x^n} m_n^{(1)}(x) A(x) P(x,t), \label{8}
\end{equation}
where $m_n^{(1)}(x)$ is the $n$th moment of $w_{1}(x \rightarrow x')$, i.e.,
$m_n^{(1)}(x)=\int (x'-x)^n w_1(x \rightarrow x') dx'$.

When we stop the expansion at the second order, as is often adopted in statistical
physics, we obtain
\begin{equation}
\bubun{P(x,t)}{t} = (A(x)-\bar{A}(t)) P(x,t) + \gamma \bubun{{}}{x}
\left[ - m_1^{(1)}(x) + \frac{1}{2} \bubun{{}}{x} m_2^{(1)}(x) \right]
A(x) P(x,t). \label{9}
\end{equation}
Eqs. (\ref{8}) and (\ref{9}) are basic equations for the evolution of distribution
function.  Eq. (\ref{9}) is an approximation. However, it is often more tractable, with
the help of techniques developed for solving the Fokker-Planck equation ( see Appendix
\ref{10} and \cite{PhysicalBiology}), while there is no established standard method for
solving eq.  (\ref{8}).

At the boundary condition we naturally impose that there are no probability flux, which is
given by
\begin{equation} \left.
\sum_{n=1}^{\infinity} \frac{(-1)^n } {n!} \bubun{{}^{(n-1)}}{x^{(n-1)}} m_n^{(1)}(x) A(x)
P(x,t) \right|_{x=x_1, x_2} =0, \label{26} \end{equation} in the case of (\ref{8}) and
\begin{equation}
\left. \left[ - m_1^{(1)}(x) + \frac{1}{2} \bubun{{}}{x} m_2^{(1)}(x) \right] A(x) P(x,t)
\right|_{x=x_1, x_2} =0
\label{27}
\end{equation}
in the case of (\ref{9}), where $x_1$ and $x_2$ are the values of the left and right
boundaries, respectively.

Next, as an example of the application of our formula, we derive the evolution equation
for Eigen's model, and estimate the error threshold, with the help of a singular
perturbation theory. Through this application, we can see the validity of eq. (\ref{9}) as
an approximation of eq. (\ref{8}).

Two additional remarks: First, introduction of the death of individuals is rather
straightforward. By including the death rate $D(x)$ into the evolution equation, the first
term in eq. (\ref{8}) (or eq. (\ref{9})) is replaced by
$\left[(A(x)-D(x))-(\bar{A}(t)-\bar{D}(t))\right] P(x,t)$, where $\bar{D}(t) \equiv \int
D(x) P(x,t) dx$. Second, instead of deriving each term in eq. (\ref{9}) from microscopic
models, it may be possible to adopt it as a phenomenological equation, with parameters (or
functions) to be determined heuristically from experiments.


\section{Application of error threshold in Eigen model}
\label{20}

In the Eigen model\cite{Eigen}, the set $S$ of the site state values is given by
$\{-1,1\}$, and the fitness (replication rate) of the sequence is given as a function of
its Hamming distance from the target sequence $\{1,...,1\}$, i.e., the fitness of an
individual sequence is given as a function of the number $n$ of the sites of the sequence
having value $1$. Hence it is appropriate to define a phenotypic value $x$ in the Eigen
model as a monotonic function of the number $n$; we determine it as $x=\frac{2n-N}{N}$, in
the range $[-1,1]$. Accordingly, the replication rate $A$ of the sequence can be written
as a function of $x$, i.e., $A(x)$; it is natural to postulate that $A$ is a non-negative
and bounded function over the whole domain. If the sequence length $N$ is sufficiently
large, the phenotypic variable $x$ can be regarded as a continuous variable, since the
step size of $x$ ($\Delta x=\frac{2}{N}$) approaches 0 as $N$ goes to infinity.

In order to derive the evolution equation of form (\ref{8}) corresponding to the Eigen
model, we only need to know the function $w_1$ in that model. (Recall that in our
formulation the mutation rate $\mu$ is assumed to be so small that only a single-point
mutation is considered.) Due to the assumption of the symmetric mutation, this
distribution function is obtained as $w_1(x \rightarrow x - \Delta x)=\frac{1+x}{2}$,
$w_1(x \rightarrow x + \Delta x)=\frac{1-x}{2}$, and $w_1(x \rightarrow x')=0$ for any
other $x'$. Accordingly, the $n$th moment is given by $m_n^{(1)}(x)= \frac{1+x}{2}
(-\Delta x)^n + \frac{1-x}{2} (\Delta x)^n$. Now, we obtain
\begin{equation}
\bubun{P(x,t)}{t} = (A(x)-\bar{A}(t)) P(x,t) + \gamma \sum_{n=1}^{\infinity} \frac{1} {n!}
\bubun{{}^n}{x^n} \left[ \frac{1+x}{2} \left( \frac{2}{N} \right)^n + \frac{1-x}{2}
\left(- \frac{2}{N} \right)^n \right] A(x) P(x,t) \label{12}
\end{equation} where
$\gamma=N \mu$, the mutation rate per sequence. When we ignore the moment terms higher
than the second order, we have
\begin{equation} \bubun{P(x,t)}{t} = (A(x)-\bar{A}(t)) P(x,t) + \frac{2
\gamma}{N} \bubun{}{x} \left[ x + \frac{1}{N}
\bubun{}{x} \right]A(x) P(x,t).
\label{11}
\end{equation}

In fact, if we focus on a change near $x\sim 0$ ( to be specific $x \sim O(1/\sqrt{N})$),
the truncation of the expansion up to the second order is validated (Or equivalently, if
we define $x'=(2n-N)/\sqrt{N}$ instead of $(2n-N)/N$, and expand eq.(3) by $1/\sqrt{N}$
instead of $1/N$, terms higher than the second order are negligible, as is also discussed
in \cite{Levine}. However, in this case, the validity is restricted to $x' \sim O(1)$
(i.e., $(n-N/2) \sim O(1)$), which means $x\sim O(1/\sqrt{N})$ in the original variable).

Now, we solve the eq. (\ref{11}) with a standard singular perturbation method (see
Appendix \ref{10}), and then return to eq. (\ref{12}).  According to the analysis in
Appendix \ref{10}, the stationary solution of the equation of form (\ref{11}) is given by
the eigenfunction corresponding to the largest eigenvalue of the linear operator $L$
defined by $L=A(x)+2 \gamma \varepsilon \bubun{}{x} \left[ x + \varepsilon \bubun{}{x}
\right]A(x)$ with $\varepsilon=\frac{1}{N}$. Now we consider the eigenvalue problem
\begin{equation} A(x) P(x) + 2 \gamma \varepsilon \bubun{}{x} \left[ x + \varepsilon
\bubun{}{x} \right]A(x) P(x) = \lambda P(x) \label{23},
\end{equation} where
$P(x) \geq 0$, with $\lambda$ to be determined.

Since $\varepsilon$ is very small (because $N$ is sufficiently large), a singular
perturbation method, the WKB approximation\cite{Morse-book}, is applied. Let us put
\begin{equation} P(x)=e^{\frac{1}{\varepsilon}\int_{x0}^{x} R(\varepsilon,x') dx'},
\label{28} \end{equation} where $x_0$ is some constant and $R$ is a
function of $\varepsilon$ and $x$, which is expanded with respect to $\varepsilon$ as
\begin{equation} R(\varepsilon,x)=R_0(x)+\varepsilon R_1(x)+\varepsilon^2 R_2(x)+...
\label{22}
\end{equation} Retaining only the zeroth order terms in $\varepsilon$ in
eq. (\ref{23}), we get \begin{equation} A(x) + 2 \gamma \left[ x R_0(x) + R_0^2(x) \right]
A(x) =\lambda,
\label{24} \end{equation} which is formally solved for $R_0$ as
$R_0^{(\pm)}(x)= \frac{-x \pm \sqrt{g(x)}}{2}$ where $g(x)= x^2+\frac{2}{\gamma}
(\frac{\lambda}{A(x)}-1)$. Hence the general solution of eq. (\ref{23}) up to the zeroth
order in $\varepsilon$ is given by $P(x)=\alpha e^{\frac{1}{\varepsilon} \int_{x_0}^{x}
R_0^{(+)}(x')dx'} +\beta e^{\frac{1}{\varepsilon} \int_{x_0}^{x} R_0^{(-)}(x')dx'} $ with
$\alpha$ and $\beta$ constants to be determined.

Now, recall the boundary conditions (\ref{27}); $P$ has to take the two branches in $R_0$
as $ P(x)=\alpha e^{\frac{1}{\varepsilon} \int_{x_b}^{x} R_0^{(+)}(x')dx'} $ for $x < x_b$
and $ P(x)= \beta e^{\frac{1}{\varepsilon} \int_{x_b}^{x} R_0^{(-)}(x')dx'} $ for $x >
x_b$, where $x_b$ is defined as the value at which $g(x)$ has the minimum value. Next,
from the continuity of $P$
at $x_b$, $\alpha=\beta$ follows, while from the
continuity of $\bubun{P}{x}$ at $x_b$, the function $g$ has to vanish at $x=x_b$. This
requirement $g(x_b)=0$ determines the value of the unknown parameter $\lambda$ as
\begin{equation}
\lambda=A(x_b) (1-\frac{\gamma}{2} {x_b}^2).
\label{18:approximated-eigenvalue}
\end{equation}
From function $P$, we find that $P$ has its peak at the point $x=x_p$, where $R_0(x)$
vanishes, i.e., at $ A(x_p)=\lambda $. Then, $P(x)$ approaches $\delta(x-x_p)$ in the
limit $\varepsilon \rightarrow +0$. These results are consistent with the requirement that
the mean replication rate in the steady state be equal to the largest eigenvalue of the
system (see Appendix \ref{10}).

The stationary solution of eq.(\ref{12}) is obtained by following the same procedure of
singular perturbation. Consider the eigenvalue problem
\begin{equation}
A(x)P(x)+\gamma \sum_{n=1}^{\infinity} \frac{1} {n!}
\bubun{{}^n}{x^n} \left[ \frac{1+x}{2} \left( 2
\varepsilon \right)^n + \frac{1-x}{2} \left(- 2
\varepsilon \right)^n \right] A(x) P(x)=\lambda
P(x). \label{25} \end{equation} By putting
$P(x)=e^{\frac{1}{\varepsilon}\int_{x0}^{x} R_0(x')
dx'}$ and taking only the zeroth order terms in
$\varepsilon$, we obtain $$A(x)+\gamma \left[
\frac{1+x}{2} \left( e^{2 R_0(x)} - 1 \right) +
\frac{1-x}{2} \left( e^{-2 R_0(x)} -1 \right) \right]
A(x) =\lambda ,$$ which gives $$
R_0^{(\pm)}(x)=\frac{1}{2} \log
\frac{1+\frac{1}{\gamma} (\frac{\lambda}{A(x)}-1) \pm
\sqrt{ \hat{g}(x)}}{1+x}$$ with $\hat{g}(x)=
(1+\frac{1}{\gamma} (\frac{\lambda}{A(x)}-1))^2-(1-x^2)
$.

By defining again the value $x=x_b$ at which $\hat{g}(x)$ takes the minimum, $P$ is
represented as $ P(x)=\alpha e^{\frac{1}{\varepsilon} \int_{x_b}^{x} R_0^{(+)}(x')dx'} $
for $x < x_b$ and $ P(x)= \beta e^{\frac{1}{\varepsilon} \int_{x_b}^{x} R_0^{(-)}(x')dx'}
$ for $x > x_b$. The continuity of $\bubun{P}{x}$ at $x=x_b$ requires $\hat{g}(x_b)=0$,
which determines the value of $\lambda$ as \begin{equation} \lambda=A(x_b) \left[1-\gamma
\left(1-\sqrt{1-{x_b}^2}\right) \right].
\label{15:more-exact-eigenvalue} \end{equation}
Again, $P(x)=\delta(x-x_p)$, in the limit $\varepsilon \rightarrow +0$, with $x_p$ given
by the condition $A(x_p)=\lambda$. When $|x_b| \ll 1$, the form
(\ref{15:more-exact-eigenvalue}) approaches eq.  (\ref{18:approximated-eigenvalue})
asymptotically.  This implies that the time evolution equation (\ref{8}), if restricted to
$|x| \ll 1$, is accurately approximated by eq.(\ref{9}) that keeps the terms only up to
the second moment.

Let us estimate the threshold mutation rate for error catastrophe. This error threshold is
defined as the critical mutation rate $\gamma^{*}$ at which the peak position $x_p$ of the
stationary distribution drops from $x_p\neq 0$ to $x_p =0$, with an increase of $\gamma$.
We use the following procedure to obtain the critical value $\gamma^{*}$.

\begin{figure}[hbtp] \begin{minipage}[t]{15cm} \begin{center}
\scalebox{ 0.45 }{\includegraphics{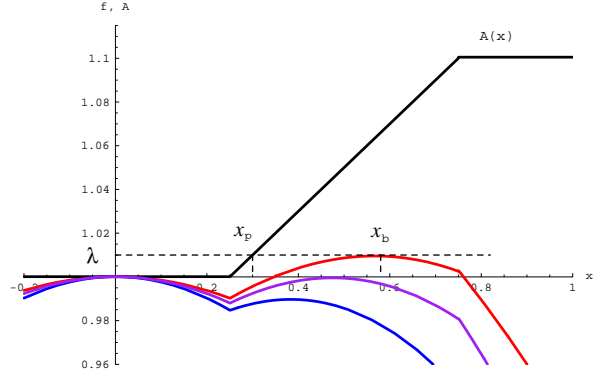}}
\caption{ Examples of profiles of the evaluation function $f$ for three values of
$\gamma$. The red, purple, and blue curves give the profiles of $f$ for $\gamma=0.31$,
$\gamma=0.386$, and $\gamma=0.49$, respectively, where $f$ is defined by $f(x)=A(x)
(1-\gamma (1-\sqrt{1-x^2}))$ and $A$ is given by $A(x)=1+0.2 (x-0.25) \Theta(x-0.25)
\Theta(0.75-x)+ 0.1 \Theta(x-0.75)$; the profile of $A$ is indicated by the black
curve. This illustrates determination of $x_b$ and $x_p$; $x_b$ is given by the position
where $f$ takes a maximum, while $x_p$ is given as the position where the line $y=f(x_b)$
crosses the curve of $A$. For $\gamma < 0.386$, $f(x)$ has a maximum value at $x=x_b$, and
thus the critical mutation rate for the error threshold is estimated to be
$\gamma^{*}=0.386$.  }
\label{33:fig:schematical-explaination-of-estimation}
\end{center} \end{minipage} \end{figure}

First consider an evaluation function whose form
corresponds to that of eigenvalue
(\ref{15:more-exact-eigenvalue}) as \begin{equation}
f(x)=A(x) \left[1-\gamma \left(1-\sqrt{1-{x}^2}\right)
\right], \label{30:more-exact-evaluation-function}
\end{equation} and find
the position at which the function $f(x)$ takes the maximum value. This procedure is
equivalent to obtaining $x_b$ in the above analysis, since the relation
$f(x)=\lambda-\frac{\gamma^2 A^2(x)}{\lambda-A(x) \left( 1-\gamma
\left(1+\sqrt{1-x^2}\right) \right)} \hat{g}(x)$ and the requirement on $x_b$ that
$\hat{g}(x_b)=0$ and $\left. \frac{d \hat{g}(x)}{dx} \right|_{x=x_b}=0$ lead to
$\left. \frac{df(x)}{dx} \right|_{x=x_b}=0$. Obviously, $x_b$ is given as a function of
$\gamma$, thus, we denote it by $x_b(\gamma)$. The position $x_b$ determines the position
$x_p$ of the stationary distribution through the relation $A(x_p)=\lambda=f(x_b)$ as in
the above analysis. If $A$ has flat parts around $x=0$ and higher parts in the region ($x
> 0$), $x_p(\gamma)$ discontinuously changes from $x_p \neq 0$ to $x_p = 0$ at some
critical mutation rate $\gamma^{*}$, when $\gamma$ increases from zero.  A schematic
illustration of this transition is given in
Fig.(\ref{33:fig:schematical-explaination-of-estimation}).

As a simple example of this estimate of error threshold, let us consider the case
\begin{equation} A(x)=1+A_0 \Theta(x-x_0),
\label{14:step} \end{equation} with $A_0>0$ and
$0<x_0<1$, and $\Theta$ as the Heaviside step function, defined as $\Theta(x)=0$ for $x <
0$ and $\Theta(x)=1$ for $x \geq 0$. According to the procedure given above, the critical
mutation rate is straightforwardly obtained as
$\gamma^{*}=\frac{A_0}{(1+A_0)\left(1-\sqrt{1-{x_0}^2}\right)}$, for $\gamma<\gamma^{*}$,
$x_p=x_0$ and for $\gamma > \gamma^{*}$, $x_p=0$.

{\sl Remark}

An exact transformation from the sequence model (Eigen model\cite{Eigen}) into a class of
Ising models\cite{Leuthausser, Baake} has recently been reported, such that the sequence
model is treated analytically with methods developed in statistical physics.  Rigorous
estimation of the error threshold for various fitness landscapes\cite{Baake2,Taiwan} and
relaxation times of species distribution have been obtained\cite{Taiwan2}. In fact, our
estimate (above) agrees with that given by their analysis.

Their method is indeed powerful when a microscopic model is prescribed in correspondence
with a spin model.  However, even if such microscopic model is not given, our formulation
with a Fokker-Planck type equation will be applicable because it only requires estimation
of moments in the fitness landscape. Alternatively, by giving a phenomenological model
describing the fitness without microscopic process, it is possible to derive the evolution
equation of population distribution. Hence, our formulation has a broad range of potential
applications.

\section{Consideration of phenotypic fluctuation}

In this section, we include the fluctuation in the mapping from genetic sequence to the
phenotype into our formula, and examine how it influences the error catastrophe. We first
explain the term ``phenotypic fluctuation'' briefly, and show that in its presence our
formulation (\ref{8}) remains valid by redefining the function $A(x)$. By applying the
formulation, we study how the introduction of the phenotypic fluctuation changes the
critical mutation rate $\gamma^{*}$ for the error catastrophe.

In general, even for individuals with identical gene sequences in a fixed environment, the
phenotypic values are distributed. Some examples are the activities of proteins
synthesized from the identical DNA \cite{Yang-et-al}, the shapes of RNA molecules of
identical sequences \cite{ancel-fontana}, and the numbers of specific proteins for
isogenic bacteria \cite{ Elowitz,Kaern-Collins,Collins,Furusawa}. Next, the phenotype $x$
from each individual with the sequence $s$ is distributed, which is denoted by
$P_{phe}(s,x)$.

We assume that the form of distribution $P_{phe}$ is characterized only in terms of its
mean value, i.e., the distributions ${P_{phe}}'s$ having the same mean value $X$ take the
same form. By representing the mean value of the phenotype $x$ by $\bar{x}(s)$, the
distribution $P_{phe}$ is written as $P_{phe}(s,x)=\hat{P}_{phe}(\bar{x}(s),x)$, where
$\hat{P}_{phe}$ is a function of $\bar{x}$ and $x$, which is normalized with respect to
$x$, i.e., satisfying $\int \hat{P}_{phe}(\bar{x},x) dx =1$.

In our formulation, the replication rate $A$ of the sequence with the phenotypic value $x$
is given by a function of phenotypic value $x$, denoted by $A(x)$.  The mean replication
rate $\hat{A}$ of the species $s$ is calculated by
\begin{equation} \hat{A}(\bar{x}(s))=\int \hat{P}_{phe}(\bar{x}(s),x) A(x)
dx. \label{phe:mean} \end{equation} 

As in the case of (\ref{1}), we assume that the transition probability from $s$ to $s'$
during the replication is represented only by its mean values $\bar{x}(s)$ and
$\bar{x}(s')$, i.e., the transition probability function is written as $W(\bar{x}(s)
\rightarrow \bar{x}(s'))$. With this setup, the population dynamics of the whole sequences
is represented in terms of the distribution of the mean value $\bar{x}$ only, so that we
can use our formulation (\ref{8}) even when the phenotypic fluctuation is taken into
account; we need only replace the replication rate $A$ in (\ref{8}) by the mean
replication rate $\hat{A}$ obtained from eq. (\ref{phe:mean}).

Now, we can study the influence of phenotypic fluctuation on the error threshold by taking
the step fitness function $A(x)$ of eq. (\ref{14:step}) and including the phenotypic
fluctuation as given in eq.(\ref{phe:mean} ).  We consider a simple case where the form of
$\hat{P}_{phe}$ is given by a constant function within a given range (we call this the
piecewise flat case). Our aim is to illustrate the effect of the phenotypic fluctuation on
the error threshold, so we evaluate the critical mutation rate $\gamma^{*}$ using the
simpler form $f(x)=A(x)(1-\frac{\gamma}{2} x^2)$ from
eq.(\ref{18:approximated-eigenvalue}), while the use of the form
(\ref{30:more-exact-evaluation-function}) gives the same qualitative result. With this
simpler evaluation function, the critical mutation rate $\gamma^{*}$ is given by
\begin{equation} \gamma_0^{*}=\frac{2 A_0}{(1+A_0) {x_0}^2},
\label{34:gamma-zero}
\end{equation} in the case without phenotypic fluctuation.
Here we examine if this critical value $\gamma^{*}_0$ increases under isogenic phenotypic
fluctuation.

We make two further technical assumptions in the following analysis: first we assume that
$A_0$ in the form (\ref{14:step}) is sufficiently small, so that the value of critical
$\gamma^{*}$ is not large. Second, we extend the range of $x$ to $[-\infinity,\infinity]$
for simplicity.  This does not cause problems because we have set the range of $x_0$ to
$(0,1)$. Hence, the stationary distribution has its peak around the range $0 \leq x < 1$;
everywhere outside this range, the distribution vanishes.

\begin{figure}[hbtp] \begin{minipage}[t]{15cm} \begin{center}
\scalebox{ 0.4 }{\includegraphics{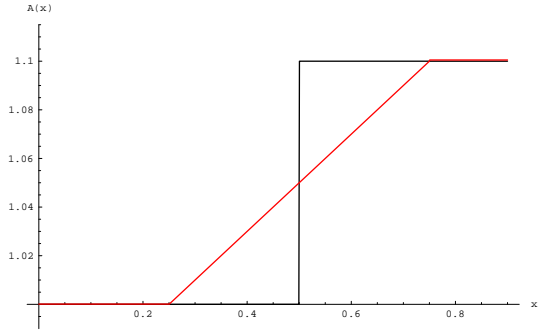}}
\caption{Example of profiles of the mean fitness functions without phenotypic fluctuation
case (black); with a constant phenotypic fluctuation over a given range given by
eq.(\ref{36:flat-case}) (red), where we set $A(x)=1+0.1 \Theta(x-0.5)$ and $\ell=0.25$.  }
\label{35:fig:profile-of-A}
\end{center} \end{minipage} \end{figure}

We consider the case in which distribution $\hat{P}_{ phe }$ of the phenotype of the
species $s$ is given by
\begin{equation}
\hat{P}_{ phe }^{(F)}(\bar{x}(s), x) = \left\{
\begin{array}{ll} 0 & \quad \mbox{for $ x
<\bar{x}-\ell$}\\ \frac{1}{2 \ell} & \quad \mbox{for $ \bar{x}-\ell \leq x \leq \bar{x} +
\ell$}\\ 0 & \quad \mbox{for $ \bar{x} + \ell < x $, }
\end{array} \right.
\label{36:flat-case}
\end{equation}
where $\ell$ gives the half-width of the distribution. ($(F)$ represents the
piecewise-flat distribution case). Then, $\hat{A}$ is calculated by
$$ \hat{A}^{(F)}(x) = \left\{ \begin{array}{ll} 1 & \quad \mbox{for $x<x_0-\ell$}\\ 1+
\frac{A_0}{2 \ell} (x-(x_0-\ell)) & \quad \mbox{for $x_0-\ell \leq x \leq x_0 + \ell$}\\
1+ A_0 & \quad \mbox{for $x_0 + \ell < x$. } \end{array} \right.$$ An example of
$\hat{A}^{(F)}(x)$ is shown in Fig. (\ref{35:fig:profile-of-A}). The evaluation function
$f$ in section (\ref{20}) is given by $ f^{(F)}(x)=\hat{A}^{(F)}(x) (1-\frac{\gamma}{2}
x^2) $.

We study the case where the position ${x_b^{*}}^{(F)} (\equiv {x_b^{(F)}}(\gamma^{*}))$ is
within the range $[x_0-\ell,x_0]$ because the profile of $\hat{A}^{(F)}$ shows that
${\gamma^{*}}^{(F)}$ is smaller than $\gamma^{*}_0$ if ${x_b^{*}}^{(F)}>x_0$.  If
$\frac{x_0}{2+A_0} \leq \ell < x_0$, the position ${x_b^{*}}^{(F)}$ is within the range
$[x_0-\ell,x_0]$.  In that case, ${\gamma^{*}}^{(F)}$ is given by ${\gamma^{*}}^{(F)}
\simeq \frac{A_0}{4 \ell (x_0-\ell)} $ to the first order of $A_0$. Comparing
${\gamma^{*}}^{(F)}$ with $\gamma^{*}_0$ in (\ref{34:gamma-zero}), we conclude that
${\gamma^{*}}^{(F)} < \gamma^{*}_0$ for $ 0 <\ell<\frac{2+\sqrt{2}}{4} x_0 $, and
${\gamma^{*}}^{(F)} > \gamma^{*}_0$ for $ \frac{2+\sqrt{2}}{4} x_0 <\ell< x_0$.  Hence,
when the half width $\ell$ of the distribution $P_{phe}$ is within the range
$(\frac{2+\sqrt{2}}{4} x_0,x_0)$, the critical mutation rate for the error catastrophe
threshold is increased.  In other words, the isogenic phenotypic fluctuation increases the
robustness of high fitness state against mutation.

We also studied the case in which $ \hat{P}_{ phe } (\bar{x}, x)$ decreases linearly
around its peak, i.e., with a triangular form.  In this case, the phenotypic fluctuation
decreases the critical mutation rate as long as $A_0$ is small, while it can increase for
sufficiently large values of $A_0$, for a certain range of the values of width of
phenotypic fluctuation.

\section{Discussion}
\label{32:discussion}

In the present paper, we have presented a general formulation to describe the evolution of
phenotype distribution.  A partial differential equation describing the temporal evolution
of phenotype distribution is presented with a self-consistently determined growth term.
Once a microscopic model is provided, each term in this evolution equation is explicitly
determined so that one can derive the evolution of phenotype distribution
straightforwardly.  This eq. (\ref{8}) is obtained as a result of Kramers-Moyal expansion,
which includes infinite order of derivatives.  However, this expansion is often summed to
a single term in the large number limit of base sequences, with the aid of singular
perturbation.

If the value of a phenotype variable $|x|$ is much smaller than unity (which is the
maximal possible value giving rise to the fittest state), the terms higher than the second
order can be neglected, so that a Fokker-Planck type equation with a self-consistent
growth term is derived.  The validity of this truncation is confirmed by putting
$x'=(2n-N)/\sqrt{N}$ and verifying that the third or higher order moment is negligible
compared with the second-order moment. Thus the equation up to its second order,
(\ref{9}), is relevant to analyzing the initial stage of evolution starting from a
low-fitness value.

As a starting point for our formalism, we adopted eq. (\ref{3}), which is called the
``coupled'' mutation-selection equation\cite{Hofbauer}.  Although it is a natural and
general choice for studying the evolution, a simpler and approximate form may be used if
the mutation rate and the selection pressure are sufficiently small.  This form given by
$\bubun{\hat{P}(s,t)}{t} = -\bar{A}(t) \hat{P}(s,t) + \sum_{s'} Q(s' \rightarrow s)
\hat{P}(s',t)$, is called the ``parallel'' mutation-selection
equation\cite{kimura1970,Akin}.  It approaches the coupled mutation-selection equation
(\ref{3}), in the limits of small mutation rate and selection pressure, as shown in
\cite{Hofbauer}. If we start from this approximate, parallel mutation-selection equation,
and follow the procedure presented in this paper, we obtain $\bubun{P(x,t)}{t} =
(A(x)-\bar{A}(t)) P(x,t) + \gamma \bubun{{}}{x} \left[ - m_1^{(1)}(x) + \frac{1}{2}
\bubun{{}}{x} m_2^{(1)}(x) \right] P(x,t)$.

In general, this equation is more tractable than eq. (\ref{9}), as the techniques
developed in Fokker-Planck equations are straightforwardly applied as discussed in
\cite{PhysicalBiology}, and it is also useful in describing of evolution.  Setting
$A(x)=x^2$ and replacing $m_1^{(1)}$ and $m_2^{(1)}$ with some constants, the equation is
reduced to that introduced by Kimura\cite{Kimura2}; while setting $A(x)=x$, $m_1^{(1)}(x)
\propto x$, and replacing $m_2^{(1)}$ with some constant derives the equation by
Levine\cite{Levine}.  Because our formalism is general, these earlier studies are derived
by approximating our evolution equation suitably.

Besides the generality, another merit of our formulation lies in its use of the phenotype
as a variable describing the distribution, rather than the fitness (as adopted by Kimura).
Whereas the phenotype is an inherent variable directly mapped from the genetic sequence,
the fitness is a function of the phenotype and environment, and strongly influenced by
environmental conditions.  The evaluation of the transition matrix by mutation in
eq.(\ref{8}) would be more complicated if we used the fitness as a variable, due to
crucial dependence of fitness values on the environmental conditions.  In the formalism by
phenotype distribution, environmental change is feasible by changing the growth term
$A(x)$ accordingly. Our formalism does include the fitness-based equation as a special
case, by setting $A(x)=x$.

Another merit in our formulation is that it easily takes isogenic phenotypic fluctuation
into account without changing the form of the equation, but only by modifying $A(x)$.  By
applying this equation, we obtained the influence of isogenic phenotype fluctuations on
error catastrophe.  The critical mutation rate for the error catastrophe increases because
of the fluctuation, in a certain case.  This implies that the fluctuation can enhance the
robustness of a high-fitness state against mutation.

In fact, the relevance of isogenic phenotypic fluctuations on evolution has been recently
proposed\cite{SatoPNAS,kaneko-book,KKFurusawaJTB}, and change in phenotypic fluctuation
through evolution has been experimentally verified\cite{Ito,SatoPNAS}.  In general,
phenotypic fluctuations and a mutation-selection process for artificial evolution have
been extensively studied recently.  The present formulation will be useful in analysing
such experimental data, as well as in elucidating the relevance of phenotypic fluctuations
to evolution.

\newpage

{\bf Figures}

\acknowledgements The authors would like to thank P. Marcq, S. Sasa, and T. Yomo for
useful discussion.


\appendix 

\section{Estimation of the transition probability in the NK model}
\label{29}

In the NK model\cite{well-written-NK,kauffmann-book}, the fitness $f$ of a sequence $s$ is
given by $$ f(s)=\frac{1}{N} \sum_{i=1}^{N} \omega_i(s) ,$$ where $\omega_i$ is the
contribution of the $i$th site to the fitness, which is a function of $s_i$ and the state
values of other $K$ sites. The function $\omega_i$ takes a value chosen uniformly from
$[0,1]$ at random. We assume that the phenotype $x$ of the sequence $s$ is given by
$x=f(s)$.

When $N \gg 1$, $K \gg 1$, and $K/N \ll 1$, the phenotype distribution of mutants of a
given sequence $s$ (whose phenotype is $x$) is characterized only by the phenotype $x$
(without the need to specify the sequence $s$). For showing this, we first examine the
one-point mutant case.

We consider the ``number of changed sites'' of sites at which $\omega's$ are changed due
to a single-point mutation. By assuming that the average number of changed sites is $K$,
the distribution of the number of changed sites $n$, denoted by $P_{site}(n)$, is
approximately given by
\begin{equation} P_{site}(n) \simeq e^{-\frac{(n-K)^2}{2 K }},
\label{1:appen:site} \end{equation} with the help of
the limiting form of binomial distribution.  Here, we have omitted the normalization
constant.

Next, we study the distribution of the difference between the phenotype $x$ of the
original sequence and the phenotype $x'$ of its one-point mutant, given the number $n$ of
changed sites of the single-point mutant.  We denote the distribution as $P_{dif \! 
f}(n;X)$, where $X=x'-x$. Here the average of $x'$ is $x(N-n)/N$, since $(N-n)$ sites are
unchanged. Thus, according to the central limit theorem, the distribution is estimated as
\begin{equation}
P_{dif \! f}(n;X) \simeq \exp \left[ {-\frac{(X+\frac{n}{N}
x)^2}{2 n \frac{\sigma^2}{N^2} }} \right] ,
\label{2:appen:diff} \end{equation} where $\sigma^2$ is the
variance of the distribution of the value of $\omega$.  This variance is estimated from
the probability distribution $P_{(s,\{\omega_i\})}(\omega)$ that the sequence $\omega$ is
generated.. Although the explicit form of $P_{(s,\{\omega_i\})}$ is hard to obtain unless
$\{\omega_i\}$ and $s$ are given, it is estimated by means of the ``most probable
distribution,'' obtained as 
follows: Find the distribution that maximizes the evaluation function $S$ (called
``entropy'') defined by $S=-\int_{0}^{1} P(\omega) \log P(\omega) d\omega$ under the
conditions $\int_{0}^{1} P(\omega) d\omega=1$ and $\int_{0}^{1} \omega P(\omega)
d\omega=x$. Accordingly the variance $\sigma^2$ may depend on $x$.

Combining these distributions (\ref{1:appen:site}) and (\ref{2:appen:diff}) gives the
distribution of $X$ without constraint on the number of changed sites:
$$ P(X) = \sum_{n=1}^{N} P_{site}(n) P_{dif \! f}(n;X) \simeq \exp \left[
{-\frac{(X+\frac{K}{N} x)^2}{2 K \frac{(\sigma^2+x^2)}{N^2}}} \right] .$$ This result
indicates that the phenotype distribution of single-point mutants from the original
sequence $s$ having the phenotype $x$ is characterized by its phenotype $x$ only; $s$ is
not necessary. Similarly, one can show that phenotype distribution of $n$-point mutants is
also characterized only by $x$. Hence, the transition probability in the NK model is
described only in terms of the phenotypes of the sequences, when $N \gg 1$, $K \gg 1$, and
$K/N \ll 1$.

\section{ Mathematical structure of the equation of form (\ref{9})} \label{10}

We first rewrite eq. (\ref{9}) as
\begin{equation} \bubun{P(x,t)}{t} = -\bar{A}(t) P(x,t) + L(x) P(x,t),
\label{17} \end{equation}
where $L$ is a linear operator, defined by $L(x)=A(x) +\bubun{{}}{x} f(x) +
\bubun{{}^2}{x^2} g(x) $ with $f(x)= - \gamma m_1^{(1)}(x) A(x)$ and $g(x)=
\frac{\gamma}{2} m_2^{(1)}(x) A(x)$.

The linear operator $L$ is transformed to an Hermite operator using variable
transformations (see below) so that $L$ is represented by a complete set of eigenfunctions
and corresponding eigenvalues, which are denoted by $\{\phi_i(x)\}$ and $\{\lambda_i\}$
($i=0,1,2,...$), respectively. Eigenvalues are real and not degenerated, so that they are
arranged as $\lambda_0 > \lambda_1 > \lambda_2 >...$.

According to \cite{PhysicalBiology}, $P(x,t)$ is expanded as
\begin{equation} P(x,t)=\sum_{i=0}^{\infinity} a_i(t) \phi_i(x), \label{13}
\end{equation}
where $a_i$ satisfies
\begin{equation} \frac{d a_i(t)}{dt}= a_i(t)
(\lambda_i-{\sum_{j=0}^{\infinity}}' a_j(t) \lambda_j). \label{19}
\end{equation}
The prime over the sum symbol indicates that the summation is taken except for those of
noncontributing eigenfunctions as defined in \cite{PhysicalBiology}.

Stationary solutions of eq. (\ref{19}) are given by $\{ a_{k}=1$ and $a_i=0$ for $i \ne k
\}$. Among these stationary solutions, only the solution $\{a_0=1$ and $a_i=0$ for $i \ne
0\}$ is stable. Hence, the eigenfunction for the largest eigenvalue (the largest
replication rate) gives the stationary distribution function. Now it is important to
obtain eigenfunctions and eigenvalues of $L$, in particular the largest eigenvalue
$\lambda_0$ and its corresponding eigenfunction $\phi_0$. Hence, we focus our attention on
the eigenvalue problem
\begin{equation} \left[ A(x) + \bubun{{}}{x} f(x) + \bubun{{}^2}{x^2} g(x)
\right] P(x) =\lambda P(x), \label{16} \end{equation} where $\lambda$ is a constant and P
is a function of $x$.

We can transform eigenvalue problem (\ref{16}) to a Schroedinger equation-type eigenvalue
problem as follows: First we introduce a new variable $y$ related to $x$ as
$y(x)=\int_{x_0}^{x} \sqrt{\frac{h}{g(x')}} dx'$ where $x_0$ and $h$ are constants. Next,
we consider a new function $\Psi(y)$ related to $P(x)$ as
$$\Psi(y)= \left. \sqrt{\frac{g(x)}{h}} e^{{\int_{y_0}^{y} \frac{\hat{f}(y')}{2 h} dy'}}
P(x) \right|_{x=x(y)} $$ where $y_0$ is some constant, $x(y)$ the inverse function of
$y(x)$, and $\hat{f}$ a function of $y$ defined by
$$\hat{f}(y)= \left. \sqrt{\frac{h}{g(x)}} (f(x)+\frac{1}{2} \frac{d
g(x)}{dx}) \right|_{x=x(y)} .$$

Using these new quantities $y$ and $\Psi$ and rewriting eigenvalue problem (\ref{16})
suitably, we get
\begin{equation} \left[ V(y) + h \bubun{{}^2}{y^2} \right] \Psi(y) =\lambda
\Psi(y), \label{21} \end{equation} where $V(y)=\hat{A}(y) + \frac{\frac{d \hat{f} (y)}{d
y}}{2} - \frac{\hat{f}^2(y)}{4 h}$ with $\hat{A}(y)=A(x(y))$.



\end{document}